# High Speed AB-Solar Sail*

**Alexander Bolonkin**
C&R, 1310 Avenue R, #F-6, Brooklyn, NY 11229, USA
T/F 718-339-4563, aBolonkin@juno.com, aBolonkin@gmail.com, http://Bolonkin.narod.ru

## Abstract

The Solar sail is a large thin film used to collect solar light pressure for moving of space apparatus. Unfortunately, the solar radiation pressure is very small about 9 $\mu$N/m$^2$ at Earth's orbit. However, the light force significantly increases up to 0.2 - 0.35 N/m$^2$ near the Sun. The author offers his research on a new revolutionary highly reflective solar sail which flyby (after special maneuver) near Sun and attains velocity up to 400 km/sec and reaching far planets of the Solar system in short time or enable flights out of Solar system. New, highly reflective sail-mirror allows avoiding the strong heating of the solar sail. It may be useful for probes close to the Sun and Mercury and Venus.

## Introduction

A solar sail is a thin film reflector that uses solar energy for propulsion. The spacecraft deploys a large, lightweight sail which reflects light from the Sun (or some other source). The radiation pressure on the sail provides thrust by reflecting photons. The solar radiation pressure is very small 6.7 Newtons per gigawatt. That equals 9.12×10$^{-6}$ N/m$^2$ at Earth's orbit (1 AU - Astronomic Unit = 150 million km) and decreases by the square of the distance from the sun. However, the solar light pressure significantly increases near sun and not far above it can reach 0.2 - 0.35 (up o.4 on Solar surface) N/m$^2$.

**Brief history**. The conventional solar sail concept was first proposed by Friedrich Zander in 1924 [1] and gradually refined over the decades. The author proposed innovations and a new design of Solar sail in 1965 [2, 3], and theory was developed in [3] -[6]. Author offers a new revolutionary solar AB-sail. Main particularity this sail is very high reflectivity which allows the AB-sail to come very close to the Sun without great heating and to attain high light force and high speed.

This innovation allows (main advantages only): 1) to achieve very high speed up 400 km/s; 2) easy to control an amount and direction of thrust without turning a gigantic sail; 3) to utilize the solar sail as a power generator (for example, electricity generator); 4) to use the solar sail for long-distance communication systems.

**Short information about Sun**. The pressure of light equals *P = 2E/c* (where *E* is energy of radiation, *c* is light speed ($c = 3\times10^8$ m/s)). The solar light energy at Earth's orbit equals 1.4 kW/m$^2$, but near a solar surface it reaches up to $64\times10^3$ kW/m$^2$ (it increases 47 thousand time!). As the result the light pressure jumps up to 0.4 N/m$^2$. The space apparatus can get significant acceleration (up to 80 m/s$^2$) and high speed up to 400 - 500 km/s.

Spectrum of Sun is presented in fig.1.

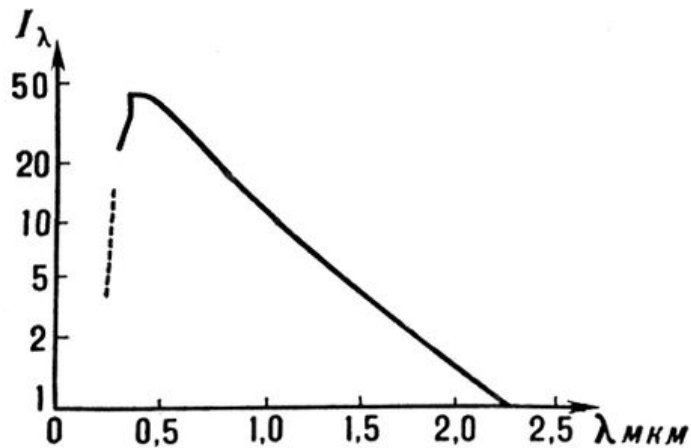

**Fig. 1.** Spectrum of solar radiation. $\lambda$ is the wavelength [0–2.5 μm], $I_\lambda$ is the energy density.

Note, the space mirror (sail) will not heat only if it reflects all solar spectrum ($\lambda = 0.2 \div 3$ μm).

## Description and Innovations

**Description.** The suggested AB space sail is presented in fig. 2. It consists of: a thin high reflection film (solar sail) supported by an inflatable ring (or other method), space apparatus connected to solar sail, a heat screen defends the apparatus from solar radiation.

The thin film includes millions of very small prisms (angle 45°, side $3 \div 30$ μm). The solar light is totally reflected back into the incident medium. This effect is called total internal reflection. Total internal reflection is used in the proposed reflector. As it is shown in [5] Ch.12 the light absorption is very small ($10^{-5} \div 10^{-7}$) and radiation heating is small (see computation section).

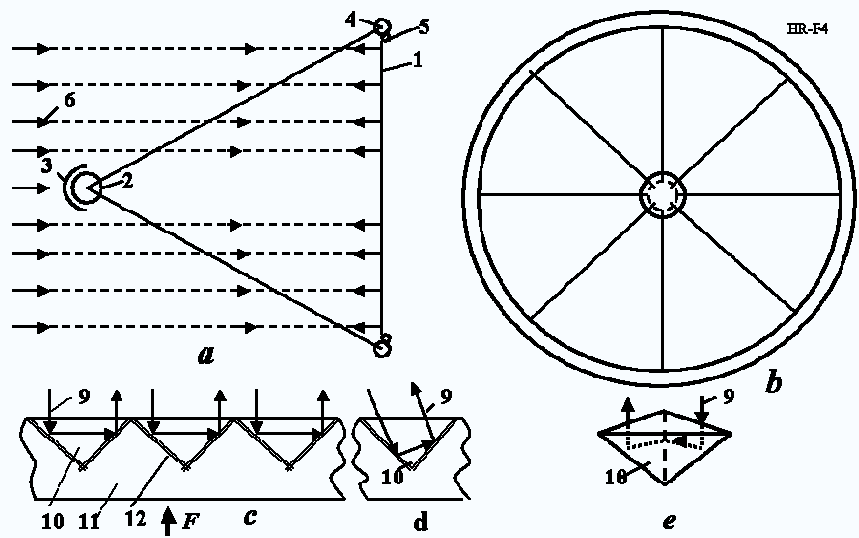

**Fig. 2**. High reflective space AB-sail. (*a*) Side view of AB-sail; (*b*) Front view; (*c*) cross-section of sail surface; (d) case of non-perpendicular solar beam; (*e*) triangle reflective sell. Notation: 1 - thin film

high reflective AB-mirror, 2 - space apparatus, 3 - high reflective heat screen (shield) of space apparatus, 4- inflatable support thin film ring, 5 - inflatable strain ring, 6 - solar light, 9 - solar beam, 10 - reflective sell, 11 - substrate, 12 - gap.

Another possible design for the suggested solar sail is presented in fig.3. Here solar sail has concave form (or that plate is made like Fresnel mirror). The sail concentrates solar light on a small control mirror 4. That mirror allows re-directed (reflected) solar beam and to change value and direction of the sail thrust without turning the large solar sail. Between thin films 1, 8 there is a small gas pressure which supports the concave form of reflector 1. Concentration of energy can reach $10^3 \div 10^4$ times, temperature greater than 5000 °K. This energy may be very large. For the sail of 200×200 m, at Earth orbit the energy is $5.6 \times 10^4$ kW. This energy may be used for apparatus propulsion or other possibilities (see [5]), for example, to generate electricity. The concave reflector may be also utilized for long-distance radio communication.

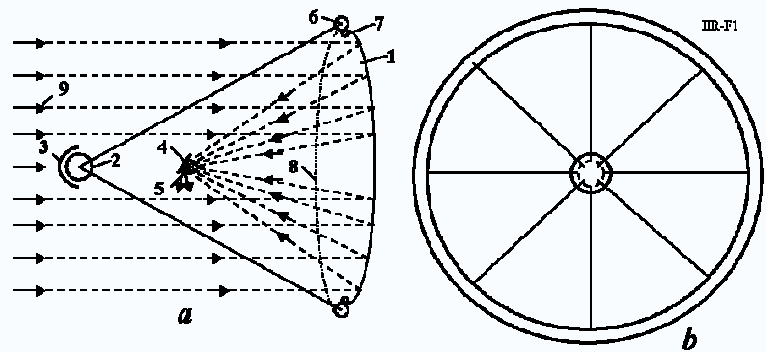

**Fig. 3**. AB highly reflective solar sail with concentrator. (a) side view; (b) front view. Notation: 1 - highly reflective AB mirror (it may have a Fresnel form), 2 - space apparatus, 3 - high reflective heat screen, 4 - control mirror, 5 - reflected solar beam, 6 - inflatable support thin film ring, 7 - inflatable strain ring, 8 - thin transparent film, 9 - solar beam.

The trajectory of the high speed solar AB-sail is shown in fig.4. The sail starts from Earth orbit. Then is accelerated by a solar light to up 11 km/s in opposed direction of Earth moving around Sun and leaves Earth gravitational field. The Earth has a speed about 29 km/s in its around Sun orbit. The sail will be have 29 -11=18 km/s. That is braked and moves to Sun (trajectory 4). Near the Sun the reflector is turned for acceleration to get a high speed (up to 400 km/s) from a powerful solar radiation. The second solar space speed is about 619 km/s. If AB sail makes three small revolutions around Sun, it can then reach speed of a 1000 km/s and leaves the Solar system with a speed about 400 km/s. Suggested highly reflective screen protects the apparatus from an excessive solar heating.

Note, the offered AB sail allows also to brake an apparatus very efficiency from high speed to low speed. If we send AB sail to another star, it can brake at that star and became a satellite of the star (or a planet of that solar system).

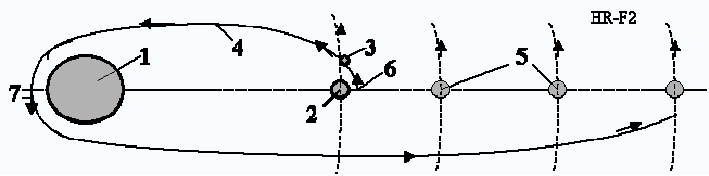

**Fig. 4**. Maneuvers of AB solar sail for reaching a high speed: braking for flyby near Sun, great acceleration from strong solar radiation and flight away to far planets or out of our Solar system. Notation: 1 - Sun, 2 - Earth, 3 AB Solar sail, 4 - trajectory of solar sail to Sun, 5 - other planets, 6, 7 - speed of solar sail.

## 3. Estimation and computation

1. **Light pressure** is calculated by equation

$$p = (1+\rho)\frac{E}{c}, \quad \text{for} \quad \rho = 1, \quad p = 2\frac{E}{c}, \qquad (1)$$

where $p$ is light pressure, N/m$^2$; $E$ is energy, J/m$^2$; $c = 3 \times 10^8$ m/s is light speed; $\rho$ is reflective coefficient ($\rho = 0 \div 1$). At solar surface $E = 64 \times 10^3$ kW/m$^2$ and $p = 0.4$ N/m$^2$. At Earth's orbit the $E = 1,4$ kW/m$^2$ and $p = 9$ μN/m$^2$.

2. **Temperature of sail** equals

$$T = 100 \sqrt[4]{\frac{\gamma E}{c_S(\varepsilon_1 + \varepsilon_2)}}, \qquad (2)$$

where $T$ is temperature, °K; $E$ is heat flow, W/m$^2$; $\gamma$ is absorption coefficient of light energy, $c_S = 5.67$ is coefficient, $0 < \varepsilon < 1$ is coefficients of blackness (emissivity) of two sail sides.

In [5] Ch. 12, Annt. #3 it is shown the absorption coefficient may reaches $\gamma = 10^{-7}$ for suggested mirror. If it is taken $\gamma = 10^{-4}$, $\varepsilon_1 = \varepsilon_2 = 0.9$, the sail temperature near the sun will be about 500 °K. That temperature is safe for many dielectric materials. The tangential sail speed in nearest point to Sun reaches 600 km/s and time of AB sail abiding near Sun is only some minutes.

3. **Trajectory and speed**. The apparatus (sail) radial speed and flight time can be estimated by equations [5] p.322.

$$V^2 = 2as_0^2\left(\frac{1}{s_0} - \frac{1}{s}\right), \quad V_{max} = \sqrt{2as_0}, \quad a = \frac{pA}{M_S + M_a}, \quad M_S = Ad, \quad t \approx \frac{s}{V_{max}}, \qquad (3)$$

where: $V$ is radial sail speed, m/s; $V_{max}$ is maximum radial sail speed, m/s; $a$ is initial (maximal) sail acceleration, m/s$^2$; $s$ is distance of the sail from a Sun center, m; $s_0$ is minimal distance, m; $p=(0.25 \div 0.4)$ is maximal light pressure [Eq.(1)], N/m$^2$; $M_S$ is mass of sail; $A$ is sail area, m$^2$; $d = (0.001 \div 0.005)$ is specific mass of sail, kg/m$^2$; $t$ is flight time from Sun to far planets, sec.

For example: If $A = 200 \times 200 = 4 \times 10^4$ m$^2$, $d = 0.005$ kg/m$^2$, $p = 0.3$ N/m$^2$, $M_a = 100$ kg, that $a = 40$ m/s.

The period of an elliptic rotation of apparatus around Sun or planet may be computed by equation

$$T_1 = \frac{2\pi}{\sqrt{K}}a_1^{3/2}, \quad K = g_0 s_0^2, \qquad (4)$$

where $T_1$ is period of rotation, sec; $a_1$ is semi-axis of big axis of ellipse, m; $g_0$ is planet (star) gravitation at distance $s_0$, m/s$^2$, (for Sun $K \approx 1.33 \times 10^{20}$ m$^3$/s$^2$; $g_0 \approx 274$ m/s$^2$; $s_0 \approx 700 \times 10^6$ m; for Earth $K \approx 4 \times 10^{14}$ m$^3$/s$^2$, $g_0 \approx 9.81$ m/s$^2$; $s_0 \approx 5.378 \times 10^6$ m ).

Computations are presented in fig. 5-7. It can be seen that the AB sail can reach very high speed (up 400 km/s) at distance 10 millions km (<1 AU) from Sun and a purview of Solar System . The flight time from Sun to the far planets is short time if we use the AB space sail (to Pluto about 150 days). But we must add a time of braking (from 29 km/s to ≈ 1 km/s) and about 65 days moving from Earth orbit to Sun (trajectory 4 in fig. 4).

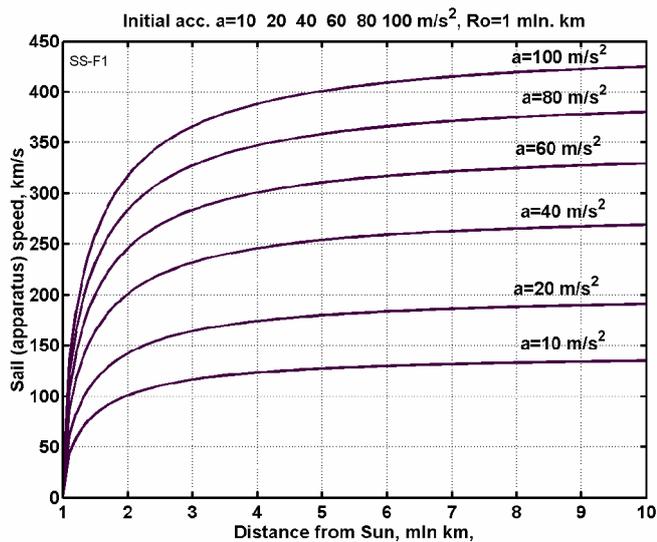

**Fig.5.** Approximately radial AB-sail speed versus distance from Sun for several initial accelerations $a$ (acceleration at minimum distance from Sun)

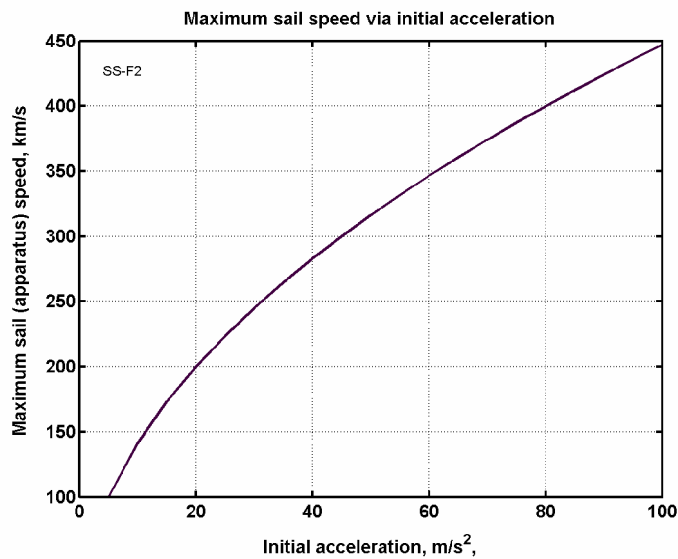

**Fig.6.** Maximal sail speed versus initial sail acceleration.

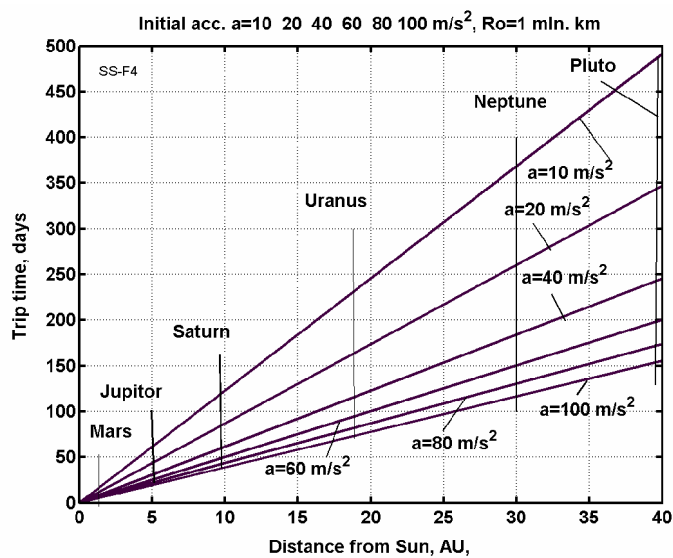

**Fig.7**. Trip time from Sun to far planets versus a distance trip from Sun.

The main particularity of offered AB-sail-reflector is special layer with very high reflectivity in the full main range of a solar spectrum (fig.1) from 0.1 to 5 μm. That means the temperature of offered sail will be significantly lower then the solar temperature and safe, allowable operating for offered layers and AB-sail-reflector.

The other particularity is special selective coating which has high thermal emissions close to absolute black body in widely range of solar spectrum.

## Discussion

The conventional mirror or multilayer dielectric mirror [12] is useless in this case. They have a high reflectivity only in narrow range of solar spectrum (Fig. 1) and decrease the adsorbed solar energy up 2 -5%. The solar surface has temperature about 5800 °K and melts any dielectric layers together with sail-mirror.

## Conclusion

The suggested new AB sail can fly very close to the Sun's surface and get high speed which is enough for quick flight to far planets and out of our Solar System. Advantages allow: 1) to get very high speed up 400 km/s; 2) easy to control an amount and direction of thrust without turning a gigantic sail; 3) to utilize of the solar sail as a power generator (for example, electricity generator); 4) to use the solar sail for long-distance communication systems.

The same researches were made by author for solar wind sail and other propulsion [7]-[11].

## Acknowledgement

The author wishes to acknowledge Richard Cathcart for correcting the English and other useful advices

## References

(Part of this article the reader can find in author WEB page: http://Bolonkin.narod.ru/p65.htm and in the book "*Non-Rocket Space Launch and Flight*", Elsevier, London, 2006,488 pgs.)